\documentclass[amsmath,amssymb,aps,pre,twocolumn,superscriptaddress]{revtex4-1}

\usepackage{graphicx}
\usepackage{braket}
\usepackage{xcolor} 

\begin{document}
\preprint{APS/123-QED}

\title{Exact evaluation of the causal spectrum and localization properties of electronic states on a scale-free network}

\author{Pinchen Xie}
\email{pinchen.xie@gmail.com}
\author{Bingjia Yang}
\thanks{Equal contribution}
\email{bjyang14@fudan.edu.cn}
\affiliation {Department of Physics, Fudan University,
Shanghai 200433, China}
\affiliation {Shanghai Key Lab of Intelligent Information
Processing, Fudan University, Shanghai 200433, China}

\author{Zhongzhi Zhang}
\email{zhangzz@fudan.edu.cn}
\affiliation {Shanghai Key Lab of Intelligent Information
Processing, Fudan University, Shanghai 200433, China}
\affiliation {School of Computer Science, Fudan University,
Shanghai 200433, China}

\author{Roberto F. S. Andrade}
\email{randrade@ufba.br}
\affiliation{Instituto de Fisica, Universidade Federal da Bahia, 40210-210, Salvador, Brazil}

\date{\today}

\begin{abstract}
A nearest-neighbor tight-binding model on a tree structure is investigated. The full energy spectrum of the normalized Hamiltonian can be expressed in terms of successively increasing number of contributions at any finite step of construction of the tree, resulting in a causal chain. The degree of quantum localization of any eigenstate, measured by the inverse participation ratio (IPR), is also analytically expressed by means of terms in corresponding eigenvalue chain. The resulting IPR scaling behavior is expressed by the tails of eigenvalue chains as well.

\end{abstract}

\maketitle

\newcommand{\G}{\mathcal{G}}
\renewcommand{\S}{\mathcal{S}}
\renewcommand{\H}{\mathcal{H}}
\newcommand{\T}{\mathcal{T}}
\renewcommand{\L}{\mathcal{L}}

\section{Introduction}
Most physical systems living on  heterogenous structure  behaves quite different from those living on homogeneous  backgrounds with translational invariance and other symmetries, as exemplified by the analysis of several quantum models in the past few decades~\cite{Alexander1982a, Sodano2006, DeOliveira2009, Jana2010, Maiti2010, DeOliveira2013}. The anomalies include, for instance, the multifractal properties of the energy spectrum, the low-dimensional Bose-Einstein condensation, and the log-periodic oscillation of thermodynamic properties ~\cite{DeOliveira2004, Coronado2005, Xie2016b}. These unique properties are usually related to exact or statistical scale invariance.

The scale-free distribution of node-degree in a large class of complex networks stands among the most interesting properties entailed by heterogeneity and, correspondingly, has received a huge attention of the community working in many different branches of science~\cite{Barabasi1999, Eguiluz2002, Barabasi2009, Radicchi2009}. Many real-world networks with the scale-free properties usually show self-similarity under a scale transformation, which leads to interesting critical behaviors for some statistical models~\cite{Cohen2001, Cohen2002b, Dorogovtsev2008, Andrade2009, DeOliveira2013, Serva2013}. On the other hand, it is possible to construct scale-free networks following an iterative self-repeating pattern~\cite{Andrade2005, Zhang2012a} according to some simple rules. The resulting deterministic network topology has a clear hierarchical structure~\cite{Lyra2014}, making it possible to evaluate the properties of physical systems with a chosen complexity using both theoretical and numerical approaches.

This study is focused on the the localization properties of non-interacting quantum system trapped on a hierarchical tree, in a tight-binding approach with only nearest neighbor interactions. The energy spectrum of tight-binding Hamiltonian models has been exactly solved for a number of self-similar scale-free networks by using renormalization technics~\cite{Serva2014, Bajorin2008a}. In most cases, the spectrum has a hierarchical structure that increases in complexity along with the iteratively constructed underlying network. The hierarchical relation among eigenvalues can be usually expressed by means of simple formulae, from which one can derive the spectrum as a one-dimension Julia set~\cite{Zhang2014, Xie2016} invariant under iteration.  For other non-scale-free networks such as the Sierpinski gasket and  Koch curve, similar hierarchical spectra also exist~\cite{Rammal1982,Andrade1989,Kappertz1994}. Along the same lines, the degree of localization of a corresponding eigenstates should depend on the hierarchical structure as well.

Despite several numerical investigations based on the use of the inverse participation ratio (IPR) to characterize the degree of quantum localization~\cite{Cardoso2008a, Mitrovic2009, Jalan2010, Goltsev2012, Pastor-Satorras2016}, we were unable to identify any contribution providing an explicit dependence between the hierarchical spectrum and corresponding eigenvector's IPR for scale-free networks. Here we consider a tight-binding model defined on an specific tree structure, for which it is possible to establish a analytical relation between the IPR (representing the degree of localization of quantum states) and the corresponding energy eigenvalue. Due to the hierarchical structure, both the eigenvalues and corresponding IPR are expressed in successive generations by means of ``causal" chains, resulting in what we call the causal spectrum of a network.

This paper is organized as follows. First we iteratively construct the  scale-free tree structure and explore its several basic properties. Then we define a heterogenous tight-binding model on the structure we construct in Sec.~\ref{cons}. Without perturbing disorder, we solve the spectrum and find its causal structure. The IPR is determined analytically given any eigenvalue chain in the spectrum.

\section{Network construction}\label{cons}
The hierarchical tree we consider here is constructed iteratively in a self-similar pattern.
\begin{figure}[hbt]
  \includegraphics[width=\linewidth]{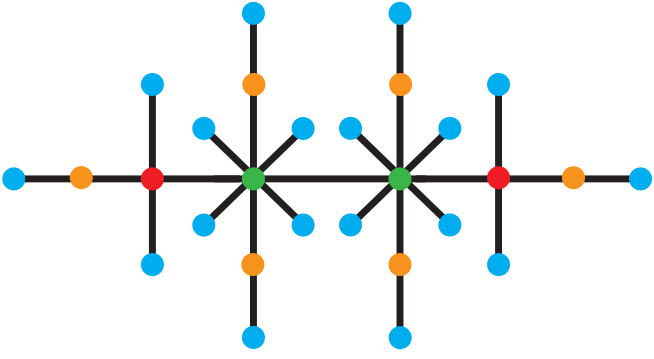}
  \caption{The construction of $G^{(3)}$. Green vertices are the initial ones. Blue vertices are created at the last generation.}\label{network}
\end{figure}

Actually, this structure corresponds to the particular case $G(m=1,\theta=0)$ of a general tree defined in~\cite{Yang2017}. Following the introduced definition and notation, $\mathcal{G}_0$ is a line with two nodes. For $t>0$, $\mathcal{G}_t$ is obtained  by attaching a  new vertex to the end of each edge in $\mathcal{G}_{t-1}$, as shown in Fig.~\ref{network}. The final infinite tree is defined as $\mathcal{G}=\lim\limits_{t\rightarrow \infty} \mathcal{G}_t$. The total number of the vertices for any finite iteration  $\mathcal{G}_t$ is
 \begin{equation}\label{N}
 	N_t=3^t+1.
 \end{equation}
A vertex introduced at the $t$-th iteration is called a  hierarchy-$t$ vertex, and their total number is precisely $2\times 3^{t-1}$. Now we label all vertices in $\mathcal{G}_t$ by $v^{(t)}_1,v^{(t)}_2,\cdots, v^{(t)}_{N_t}$, and define the adjacency matrix $A^{(t)}$ with elements $a^{(t)}_{ij}=1$ for adjacent $v^{(t)}_i$ and $v^{(t)}_j$, and $a^{(t)}_{ij}=0$ otherwise. The degree of any vertex $v^{(t)}_{i}$ in $\mathcal{G}_t$ is $d^{(t)}_i=\sum_j a^{(t)}_{ij}$. The degree of any hierarchy-$n$ vertex in $\mathcal{G}_t$ is  $d^{(t,n)}=2^{t-n}$.  Thus $G$ is scale-free with the degree distribution $P(k)\sim k^{-\gamma}$ that $\gamma=1+\frac{\ln 3}{\ln 2}$. Let us also define the diagonal degree matrix $D^{(t)}$ with elements $\delta_{ij} d^{(t)}_{i}$.

The normalized stochastic matrix~\cite{Chen2007} for Markov chains on $\mathcal{G}_t$ is $P^{(t)}=(D^{(t)})^{-\frac{1}{2}}A^{(t)} (D^{(t)})^{-\frac{1}{2}}$. Obviously, the $i,j$-th entry of $P^{(t)}$ is $p^{(t)}_{ij}=\frac{a^{(t)}_{ij}}{\sqrt{d^{(t)}_id^{(t)}_j}}$. Again, we define $P=\lim\limits_{t\rightarrow\infty} P^{(t)}$.

\section{Tight binding models on $\mathcal{G}$}\label{tb}
A typical heterogenous tight-binding model on $\mathcal{G}_t$ for non-interacting electrons is written as
\begin{equation}
	\tilde{H}_t=\sum_i d^{(t)}_i c_i^{\dagger}c_i+\sum_{i\sim j} a^{(t)}_{ij}c_i^{\dagger}c_j,
\end{equation}
where the first sum is taken over all vertices and the second sum is taken over all nearest neighbors of the network. $c_i^{\dagger}$ and $c_i$ are the electron creation and annihilation operators on site $i$. Note that the eigen-energy of any vertex is not constant, but it increases with the node degree.
The spectrum of $\tilde{H}_t$ is  unbound in the large-$t$ limit. However, by rescaling the on-site potential and the hopping amplitude at the same time, we can always renormalize the Hamiltonian to
\begin{equation}\label{hamiltonian}
	H_t=\sum_{i\sim j} p^{(t)}_{ij}c_i^{\dagger}c_j,
\end{equation}
where we have omitted the  constant on-site term $\sum_i  c_i^{\dagger}c_i$.
The spectrum of $H_t$ is bounded, lying on the $[-1,1]$ interval~\cite{Huang2015}.

Let $\varepsilon$ be any eigenvalue of $H_t$ and $\Phi^{(t)}=\sum_i \phi^{(t)}_i \ket{i}$ the corresponding eigenstate, where $\ket{i}$ is the totally localized state at vertex $i$.
The inverse participation ratios (IPR) of a state $\Phi^{(t)}$ is defined as
${\xi }=1/\sum_i (\phi^{(t)}_i)^4$. If we allow $\Phi^{(t)}$ to be  unnormalized, ${\xi}$ is redefined as
\begin{equation}\label{ipr}
	{\xi }=\frac{\left(\sum_i (\phi^{(t)}_i)^2\right)^2}{\sum_i (\phi^{(t)}_i)^4}.
\end{equation}
${\xi}$ is a measure of the effective number of vertices covered by a state, taking values in the interval $[1,N]$, where the extremes correspond, respectively, to the completely localized and completely extended states.

Now suppose $i$ is any hierarchy-$n$ vertex $(n<t)$ in $\mathcal{G}_t$. Note that it has $2^{t-n-1}$ hierarchy-$t$ neighbors and $2^{t-n-1}$ lower hierarchy neighbors. Since $p^{(t)}_{ij}={a^{(t)}_{ij}}/\sqrt{d^{(t,n)}}$ for any vertex $j$ of hierarchy-$t$ and  $p^{(t)}_{ik}={a^{(t)}_{ik}}/{\sqrt{d^{(t,n)}d^{(t,m)}}}$ for any hierarchy-$m$ vertex $k$ with $(m<t)$, we can write the condition satisfied by any eigenstate $\Phi_t$
\begin{equation}\label{0}
	\varepsilon \phi^{(t)}_i=\sum_{j=1}^{N_t} p^{(t)}_{ij}\phi^{(t)}_j
\end{equation}
by partitioning the above sum over the index $j$ into two parts
\begin{equation}\label{1}
	\varepsilon \phi^{(t)}_i=\sum_{j}\frac{1}{\sqrt{d^{(t,n)}}}\phi^{(t)}_j+\sum_{k}  \frac{a^{(t)}_{ik}}{\sqrt{d^{(t,n)}d^{(t,m)}}}  \phi^{(t)}_k.
\end{equation}
The second sum is taken over all neighbors of node $i$ with hierarchy $m<t$.

On the other hand, any hierarchy-$t$ vertex has just one hierarchy-$n$ neighbor with $n<t$. Therefore, Eq.(\ref{0}) becomes
 \begin{equation}\label{2}
 \varepsilon \phi^{(t)}_j=\frac{1}{\sqrt{d^{(t,n)}}}\phi^{(t)}_i.
 \end{equation}
After inserting Eq.~(\ref{2}) into Eq.~(\ref{1}) we obtain
\begin{equation}\label{3}
	\varepsilon \phi^{(t)}_i=\sum_{j}\frac{1}{\varepsilon {d^{(t,n)}}}\phi^{(t)}_i+\sum_{k}  \frac{a^{(t)}_{ik}}{\sqrt{d^{(t,n)}d^{(t,m)}}}  \phi^{(t)}_k,
\end{equation}
where the first sum has $2^{t-n-1}$ terms as mentioned before. Taking into account that $d^{(t,n)}=2d^{(t-1,n)}$, Eq.~(\ref{3}) can be rewritten in a form resembling the eigenstate equation for the preceding generation $\mathcal{G}_{t-1}$ as
\begin{equation}\label{iter}
	\Big(2\varepsilon-\frac{1}{ \varepsilon }\Big) \phi^{(t)}_i= \sum_{k}\frac{a^{(t)}_{ik}}{\sqrt{d^{(t-1,n)}d^{(t-1,m)}}}  \phi^{(t)}_k.
\end{equation}

\begin{figure}[hbt]
\begin{center}
		\includegraphics[width=\linewidth]{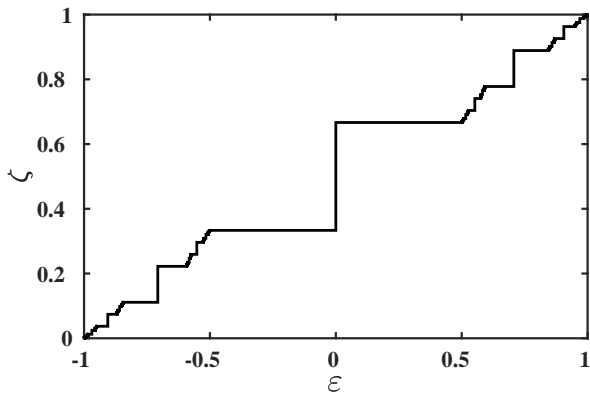}
		  \caption{Integrated density of states $\zeta$ as a function of the eigenvalue $\varepsilon$.}\label{IDOS}
\end{center}
\end{figure}

Eq.~(\ref{iter}) specifies an exact coarse-graining transition from $\mathcal{G}_t$ to $\mathcal{G}_{t-1}$. Consequently, the discrete flow equation ($\varepsilon\neq 0 $)
\begin{equation}\label{R}
	\varepsilon'={\cal S}(\varepsilon )=2\varepsilon-\frac{1} {\varepsilon }
\end{equation}
yields the eigenvalue $\varepsilon'$ of the Hamiltonian $H_{t-1}$ for $\mathcal{G}_{t-1}$. A more systematic treatment of the flow equation related to a general tree can be found at \cite{Yang2017}.

\begin{figure}[bt]
\includegraphics[width=0.7\linewidth]{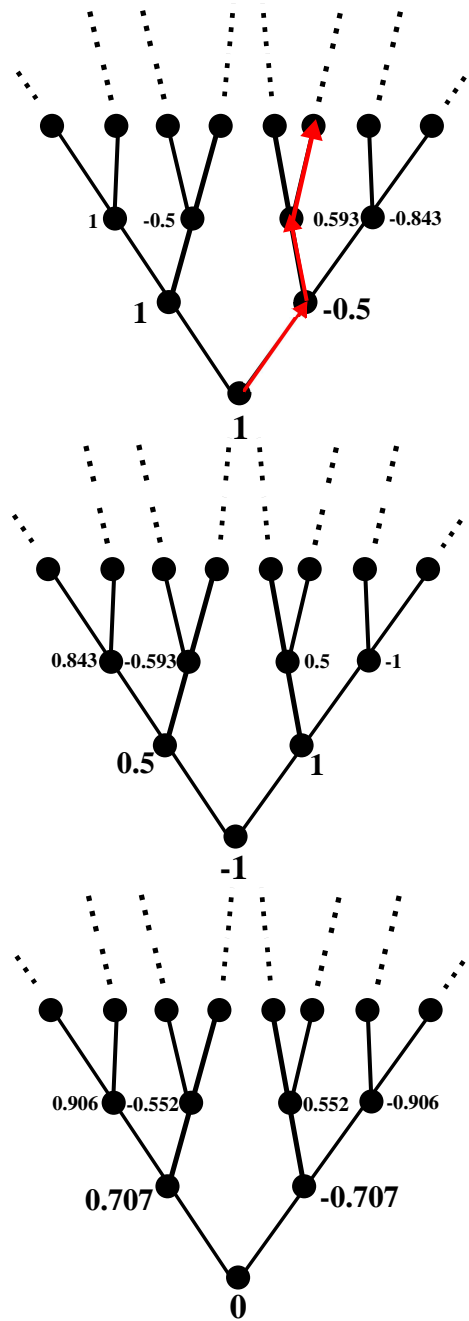}
\caption{The schematic representation of the causal spectrum. Eigenvalues $1$ and $-1$ are eigenvalues related to the original network $\mathcal{G}_0$. $0$, the singularity of $\cal S$, is  an eigenvalue with huge degeneracy in the large-$t$ limit. Other nodes originate from $1$, $-1$ or $0$ by taking the inverse of $\cal S$. The arrowed lines in the first lattice is an example of eigenvalue chains.}\label{spectrum}
\end{figure}

Note that ${\cal S}(\varepsilon)$ provides only one half of the hierarchical structure of the full spectrum.  The other half is induced by the singularity of ${\cal S}(\varepsilon)$ at $\varepsilon=0$.  Unlike other eigenvalues, the $\varepsilon=0$ eigenvalue of $\mathcal{G}_t$ has no correspondence to any eigenvalue of $\mathcal{G}_{t-1}$, which makes $\varepsilon=0$ itself the origin of a hierarchical tree of eigenvalues for any value of $t$. Besides that, this eigenvalue has a huge degeneracy, as illustrated by the schematic representation of the integrated density of states of the spectrum in Fig.~\ref{IDOS}. Similar results on hierarchical spectrum can be found at~\cite{Teplyaev1998,Zhang2009a,Xie2016b}. To show the full hierarchical structure of the entire spectrum of  $\mathcal{G}_t$, we draw the lattice-like representation of the spectrum in Fig.~\ref{spectrum}. From below to the top, we define the ``causal" relation using directional edges against the direction of renormalization. An illustrative eigenvalue  chain  is also indicated in the figure. By comparing Figs.~\ref{IDOS} and ~\ref{spectrum} it is also possible to identify that huge degeneracy and wide gaps appear for all eigenvalues that emerge from $\varepsilon=0$.

We observe that Eq.~(\ref{iter}) also gives the information about the eigenstate associated to ${\cal S}(\varepsilon )$ for $\mathcal{G}_{t-1}$, once we can write $\Phi^{(t-1)}=\sum_{i\in \mathcal{G}_{t-1} } \phi^{(t-1)}_i\ket{i}=\sum_{i\in \mathcal{G}_{t-1} } \phi^{(t)}_i\ket{i}$. Obviously, $\Phi^{(t-1)}$ is merely a segmentation of $\Phi^{(t)}$ and we intentionally ignore the normalization condition.

We would like to remind that several aspects of the derived results, e.g., the direct and inverse recurrence relations between eigenvalues of successive values of $t$, the possibility of expressing their values according to the tail of a branching process in the spectrum, the presence of high degenerate eigenvalues in the spectrum, have been used in the study of linear models with deterministic disorder ~\cite{Schellnhuber1985,Andrade1989,Kappertz1994}. In such cases, other analytical and numerical approaches to identify the extended/critical/localized nature of the quantum states have been used, as the band-gap ratio associated to the rotation number that characterizes the energy spectrum. The analytical approach for the evaluation of $\xi$  we present in the next section advances beyond its usual  numerical evaluation in terms of eigenvector components.

\section{Analytical evaluation of $\xi$}

\subsection{Recurrence relations}

The next step we obtain the analytical expression relating $\xi$ to the eigenvalue chains. If we raise both sides of Eq.~(\ref{2}) to the power $r$, we obtain
\begin{equation}\label{xi}
	\varepsilon^r (\phi^{(t)}_j)^r =(d^{(t,n)})^{\frac{-r}{2}}(\phi^{(t)}_i)^r=2^{-(t-n)r/2}(\phi^{(t)}_i)^r.
\end{equation}
Define $\Theta_n(r)$ ($n\leq t$) as the sum of  $(\phi^{(t)}_i)^r$ over all vertices of hierarchy-$n$.  Since we keep all eigenstates unnormalized, $\Theta_n(r)$ is independent of $t$.  Sum Eq.~(\ref{xi}) over all vertices of hierarchy-$t$:
 \begin{equation}
 	\varepsilon^r\Theta_t(r)=\sum_{n=0}^{t-1}  2^{(2-r)(t-n)/2-1}\Theta_n(r).
 \end{equation}
 Here $\varepsilon$ belongs to $\mathcal{G}_t$. For $\mathcal{G}_{t-m}$, the consistency of $\Theta_n(r)$ requires \begin{equation}\label{chain}
 		({\cal S}^{\otimes m}(\varepsilon))^r\Theta_{t-m}(r)=\sum_{n=0}^{t-m-1}  2^{(2-r)(t-m-n)/2-1}\Theta_n(r)
 \end{equation}
 where ${\cal S}^{\otimes m}(x)={\cal S}\circ\cdots\circ {\cal S}(x)$ with $m$ symbols ${\cal S}$ denotes the $m$-composite of the function ${\cal S}$, so that $({\cal S}^{\otimes m}(\varepsilon))$ indicates an eigenvalue of the $(t-m)$-th generation.

At the left hand side of Eq.~(\ref{chain}) we identify the element of a finite eigenvalue chain ending at $\varepsilon$. Let  $\Gamma^{(t)}_\epsilon$ denote this chain for any eigenvalue $\epsilon$ of $H_t$. To proceed with the analysis of Eq.~(\ref{chain}), we must consider two different circumstances.

First, for $\epsilon$ originated from one of the $t=0$ eigenvalue $\pm1$, the  length of $\Gamma^{(t)}_\epsilon$ is  always $t$. Let us relabel the chain elements from the top as $\lambda_t=\varepsilon, \lambda_{t-1}={\cal S}(\varepsilon),\cdots,\lambda_0={\cal S}^{\otimes t}(\varepsilon)$.
From  Eq.~(\ref{chain}), one obtains by elementary algebra that
\begin{equation}\label{result1}
	\frac{\Theta_{n}(r)}{\Theta_{0}(r)}=\Big(\frac{1}{2}\Big)^{\frac{r-2}{2}n} \bigg[\frac{1}{2\lambda_n^r} \prod_{k=1}^{n-1}\Big(1+\frac{1}{2\lambda_k^r}\Big)	\bigg].
\end{equation}

\noindent Hence the IPR defined by Eq.~(\ref{ipr}) can be analytically expressed by
\begin{equation}\label{result2}
 \xi=\frac{(\Theta_{0}(2))^2}{\Theta_{0}(4)}\frac{\bigg[1+\sum\limits_{n=1}^{t} \frac{1}{2\lambda_n^2} \prod\limits_{k=1}^{n-1}\Big(1+\frac{1}{2\lambda_k^2}\Big)	 \bigg]^2}{1+\sum\limits_{n=1}^{t} \bigg[\frac{1}{2^{n+1}\lambda_n^4} \prod\limits_{k=1}^{n-1}\Big(1+\frac{1}{2\lambda_k^4}\Big)	\bigg]}.
\end{equation}
Eq.~(\ref{result2}) suggests $\xi$ is fully decided by the causal chain $\Gamma^{(t)}_\epsilon$.   Also, by Eq.~(\ref{result2}) we can compute the IPR with high efficiency for any given $\varepsilon$.

However, for any $\varepsilon$ originating from $\varepsilon=0$, the above expression for $\xi$ does not hold. For the unperturbed system expressed by Eq. (\ref{hamiltonian}), $\varepsilon=0$ and all of its son eigenvalues possess huge degeneracy, which can be greatly reduced if small disorder is added to the system. Nevertheless, we can still analyze some aspects that emerge under this condition of extreme degeneracy.

For $\varepsilon=0$,  $H_t\Phi^{(t)}=0$ requires that any eigenstate $\Phi^{(t)}$ associated to $0$  vanishes at all internal sites.  However the number of degrees of freedom of $\Phi^{(t)}$ is smaller than the number of the boundary sites, because the values of $\Phi^{(t)}$ on the boundary sites are still restricted by $H_t\Phi^{(t)}=0$.  The IPR for states on this level ranges from $1$ to $4 \times 3^{t-2}$. In the large $t$ limit, $\frac{\xi}{N}\in [0,\frac{2}{3}]$.

For $\varepsilon$ originating from $0$, the length $q$ of the causal chain $\Gamma^{(t)}_\epsilon$is less than $t$. Let $q$ be the smallest integer satisfying  ${\cal S}^{\otimes q}(\varepsilon)=0$. We still label the chain element from the bottom: $\lambda_{q}=\varepsilon,\cdots, \lambda_{0}={\cal S}^{\otimes q}(\varepsilon)=0$.
For this condition, Eq.~(\ref{chain}) ends at
\begin{equation}
\lambda_1^r\Theta_{t-q+1}(r)= 2^{-r/2}\Theta_{t-q}(r)
\end{equation}
where $\lambda_1$ is either $\sqrt{1/2}$ or $-\sqrt{1/2}$ (the preimages of $0$) depending on $\varepsilon$. Notice that $\Theta_{n}(r)=0$ ($n<t-q$) holds for  eigenstates associated to $0$ on $\mathcal{G}_{t-q}$. Thus, analogously to Eq.~(\ref{result1}),   $\Theta_{n}(r)$ writes
\begin{equation}\label{result4}
	 \frac{\Theta_{n}(r)}{\Theta_{p}(r)}=\Big(\frac{1}{2}\Big)^{\frac{r-2}{2}(n-p)} \bigg[\frac{1}{2\lambda_{n-p}^r} \prod_{k=1}^{n-p-1}\Big(1+\frac{1}{2\lambda_k^r}\Big)	\bigg]
\end{equation}
where we define $p=t-q$ to make the equation neat. $\Theta_{p}(r)$ has one free degree of freedom.

Eq.~(\ref{result4}) has nearly the same form as Eq.~(\ref{result1}).
We should wrap the coefficient $\frac{(\Theta_{0}(2))^2}{\Theta_{0}(4)}$ in a  constant $C$ in Eq.~(\ref{result2}). $C$ corresponds a free degree of freedom when the renormalization flow terminates early.  Then Eq.~(\ref{result2}) will apply to both situations.

\subsection{Scaling theory}

For an infinitely long eigenvalue chain, it is necessary to study its scaling behaviors. According to Eq.~(\ref{result2}), the scaling of $\xi$ can be determined completely by the tail of a chain.

If we are trying to derive an infinite chain on one of the three lattices in Fig.~\ref{spectrum}, we should climb against the renormalization flow by taking the inverse of ${\cal S}$. For any node $\varepsilon$ in the lattices, its preimage under ${\cal S}$ resulting from the inversion of Eq.(\ref{R}) can be represented by  $\{{\cal L}(\varepsilon), {\cal R}(\varepsilon)\}$ where ${\cal L}(x)=(x+\sqrt{x^2+8})/4$ and ${\cal R}(x)=(x-\sqrt{x^2+8})/4$. Since there are only two directions to extend a chain, we can encode a chain by  a binary sequence consisting of $L$ and $R$. And we order the ``L''s and ``R''s with respect to the direction we choose at each step. For example, the chain $\Gamma=\{1,{\cal L}(1),{\cal R}({\cal L}(1)),{\cal L}({\cal R}({\cal L}(1))),\cdots \}$ is encoded as ``LRLRL$\cdots$''. It it not hard to verify that if the binary sequence of a chain's tail is periodic, the  eigenvalues at the tail have the same period as the binary sequence.  In the last example, the  repeating unit of the encoded chain is ``LR", the corresponding minimal repeating unit of the eigenvalues are simply $\{\sqrt{1/3},-\sqrt{1/3}\}$. We observe that the binary encoding of the chain follows the same strategy as that of the 4-adic expression of the rotation number characterizing the eigenvalues and localized character of the states in the Koch fractal~\cite{Andrade1989}

Starting with the simplest periodic tails of period 1, we obtain that the chain with tail ``LLL$\cdots$" converges to $1$ and a chain with tail  ``RRR$\cdots$'' converges to $-1$ instead. This is simply because $1$ and $-1$ are the only two fixed point of ${\cal S}$.
For both cases, the right hand side of Eq.~(\ref{result1}) can be approximated by $\Theta_{n}(r)=A\times 3^{n-1}/2^{rn/2}$ ($A$ is a constant) in large-$n$ limit when $r$ is even. It is easily seen that ${\sum_{n=0}^{\infty}\Theta_n(4)}$ converges and that, in large-$t$ limit, the IPR can be approximated by
\begin{equation}\label{scale1}
	\xi=\frac{(\sum_{n=0}^{t}\Theta_n(2))^2} {\sum_{n=0}^{\infty}\Theta_n(4)}=B \times \big(\frac{3}{2}\big)^{2t},
\end{equation}
where $B$ is a constant.

Recalling that $N_t=3^t+1$, we immediately obtain the scaling behavior of $\xi$ with respect to the network size $N$:
\begin{equation}\label{constant_tail}
\xi\propto N^{2(1-\log 2/\log 3)}.
\end{equation}

Using similar strategies, the scaling behavior of any given infinite chains $\Gamma$ with a periodic tail is solvable.
Indeed, let $l$ be the period of the tail. Let $\{\epsilon_1,\cdots,\epsilon_l\}$ be the minimal repeating unit of the eigenvalue chains at the tail and define
\begin{equation}
\tau_r=\Big(\prod_{i=1}^l\big(1+\tfrac{1}{2\epsilon_i^r}\big)\Big)^{\frac{1}{l}}.
\end{equation}
From Eq.~(\ref{result1}) one obtains the scaling of $\Theta_n(r)$:
\begin{equation}
\Theta_n(r)\propto\Big(\frac{1}{2}\Big)^{\frac{r-2}{2}n}\tau^n_r.
\end{equation}
Hence $\Theta_n(2)\propto \tau_2^n $ and $\Theta_n(4)\propto(\tau_4/2)^n$. In contrary to the result in Eq.(\ref{scale1}) for the simplest $l=1$ tail, the behavior of $\tau_4$, and consequently of $\Theta_n(4)$, depends on $l$ as well as on the chosen sequence of symbols. Therefore, we can not guarantee that $\Theta_n(4)$ vanishes in the large-$n$ limit for all possible periodic chains. To better analyze the consequences we assume that, in the large-$t$ limit, $\xi\propto N^\alpha$. When $\tau_4> 2$, the IPR in large-$t$ limit behaves as
\begin{equation}
	\xi\propto \frac{\tau_2^{2t}}{(\tau_4/2)^t}.
\end{equation}
Thus
\begin{equation}\label{g1}
\alpha=(2\log \tau_2 +\log 2-\log \tau_4)/\log 3.
\end{equation}

On the other hand, when $\tau_4\leq 2$, $\sum_{n=0}^{t}\Theta_n(4)$ does not grow exponentially with respect to $t$. So
\begin{equation}
\xi\propto \tau_2^{2t},
\end{equation}
which suggests
\begin{equation}\label{g2}
	\alpha=2\log\tau_2/\log 3.
\end{equation}

\begin{figure}[hbt]
  \includegraphics[width=\linewidth]{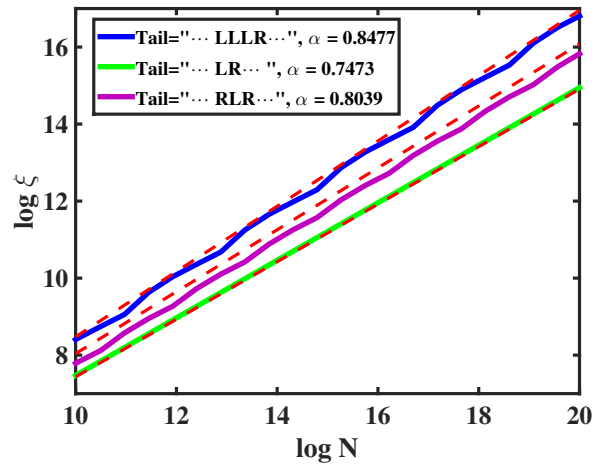}
  \caption{The scaling behaviors of three eigenvalue chains with periodic tails. The  gradients $\alpha$ of the dashed asymptotic  lines are predicted by Eq.~(\ref{g1}) and  Eq.~(\ref{g2}). The inset shows the scaling behavior of the aperiodic eigenvalue chain encoded by the Fibonacci words.}\label{3tails}
\end{figure}

To verify the accuracy of our calculations, we considered a number of periodic tails and numerically computed the IPR as a function of $t$. Fig.~\ref{3tails} shows the scaling behavior with respect to the network size for binary sequences with the following repeating units: ``LR'', ``RLR'', and  ``LLLR''. We can observe a very good agreement with the predicted analytical results.
Of course the use of Eq. (\ref{result2}) is not restricted to periodic tails. For any aperiodic chain, be it randomly or deterministically generated, the value of $\xi$ can be obtained from Eq. (\ref{result2}) by inserting the corresponding values of $\lambda_k$ in the sequence. The inset in Fig.~\ref{3tails} illustrates the kind of we obtain for an eigenvalue chain obtained by the Fibonacci sequence. 

All sequences that we investigated have systematically lead to $\alpha\ge 2(1-\log2/\log3)$, what hints that the non-degenerate states and finitely-degenerated states of the model are nearly extended. 

We investigated whether it is possible to provide a unifying relationship between the value of $\alpha$ and the period of the tail. This is a complex task once, when the period increases, the number of possible units increases exponentially. After the analysis of a large number of periodic tails, we found that the $\alpha$ depends not only on the number of symbols in the sequence, but on each specific selection of symbols. The emerging landscape is very complex, and we were unable to identify a simple rule that is valid for all sequences. Nevertheless, some trends have been identified: i) for a fixed period, tails that oscillate with high frequency between ``L'' and  ``R'' usually have lower $\alpha$, e.g., as it follows from the comparison of two period 5 sequences $\alpha($``LRLRL"$) = 0.781$ and $\alpha($``LLLLR"$)= 0.8768$; ii) for tails formed by the intercalation of one single opposite symbol amid a sequence of equal symbols, $\alpha$ increases with the number of equal symbols, as one can note by the quoted values  for  ``LR'', ``RLR'', ``LLLR'', and  ``LLLLR''. However, this behavior does not persist forever. We found that, when the number of the repeating symbols increases beyond a threshold,  $\alpha$ reverts this monotonic behavior.

Another important issue to address is the scaling of the IPRs related to a convergent eigenvalue chain instead of a periodic (oscillating) one.
Consider a finite size structure $\mathcal{G}_t$, a given eigenvalue $\varepsilon$ originating from $\pm1$ is  at the top of its eigenvalue chain $\Gamma^{(t)}_\epsilon$. Since $\Gamma^{(t)}_\epsilon$ is uniquely determined by $\epsilon$, we obtain a convergent chain by taking the limit $\Gamma_\epsilon=\lim_{t\rightarrow\infty}\Gamma^{(t)}_\epsilon$.

We realize that, when $t$ is finite but large enough, after a limited number of coarse-graining steps towards $\mathcal{G}_t$, $\varepsilon$ is mapped into $1$ or $-1$. This means that the head of the chain $\Gamma^{(t)}_\epsilon$ is constantly $1$ or $-1$. For $t\rightarrow\infty$, $\Gamma_\epsilon$ begins with an infinitely long constant sequence and ends by a certain finite tail. If we reverse $\Gamma_\epsilon$, the outcome is merely an infinite chain with a periodic tail that is actually constant. Thus, Eq.~\ref{constant_tail} directly gives the scaling exponent of the IPR of the eigenstate associated to a convergent chain without requiring any further computation.

However, for a given degenerate level $\varepsilon$ originating from $0$, it is not advisable to use the analytic expressions for the scaling behavior of $\xi$, because the degeneracy of $\varepsilon$ depends on $t$. Also, the length of the chain beneath $\varepsilon$ is constant when $t$ is large enough. Therefore, it is better to study the scaling behavior of $\xi$ for this kind of states directly from Eq. (\ref{result2}) instead of relying on scaling expressions.

\section{Conclusion}\label{conclu}
In this work, we used an exact renormalization approach to solve the energy spectrum of a network Hamiltonian with tight-binding approximation. The causal structure of the spectrum is revealed by the directed eigenvalue chains.  We found the explicit dependence between the IPR of a state and the eigenvalue chain beneath it. The scaling behavior of IPRs related to non-degenerate energy levels are  determined for non-degenerate eigenvalues and chains with periodic tails.

Our analytical study well establishes a bridge between the eigenvalue spectrum and IPR. One would expect there will be other features of scale-free networks  relying on the causal structure of a spectrum. Further studies can pay attention to finding these connections.

\begin{acknowledgments}
This work was supported by the National Natural Science Foundation of China under Grant No. 11275049 and by the Brazilian agency Conselho Nacional de Desenvolvimento Cient\'{i}fico e Tecnol\'{o}gico (CNPq) under Grant No. 310764/2016-5. P. Xie and B. Yang are also supported by the Chun-Tsung Endowment. R. F. S. Andrade benefitted from the support of the Instituto Nacional de Ci\^{e}ncia e Tecnologia para Sistemas Complexos, Brazil

\end{acknowledgments}


\bibliography{reference}

\end{document}